\documentclass[preprint,10pt]{elsarticle}

\usepackage{amsmath, amsthm, amssymb,amssymb,graphicx,fancyhdr,fancyhdr,helvet,ifpdf,setspace,graphics}
\usepackage{booktabs}
\usepackage{pstricks}
\usepackage{cancel}
\usepackage{float}
\usepackage{fullpage}
\usepackage{enumerate}
\usepackage{setspace} 
\journal{}

\begin{document}
\begin{frontmatter}

\title{A novel two dimensional nonlinear tuned mass damper inerter and its application in vibration mitigation of wind turbine blade}

\author[UNR]{V. Jahangiri}
\author[lsucee]{C. Sun\corref{cor1}}
\ead{csun@lsu.edu}
\cortext[cor1]{Corresponding author}

\address[UNR]{Department of Civil \& Environmental Engineering, University of Nevada, Reno, Nevada, USA}
\address[lsucee]{Department of Civil and Environmental Engineering, Louisiana State University, Baton Rouge, Louisiana 70803, USA}

\begin{abstract}
The present paper proposes a novel two-dimensional non-linear tuned mass damper inerter (2d-NTMDI) to effectively mitigate bi-directional structural responses. The 2d-NTMDI consists of a mass and two sets of springs, dashpots and inerters configured in two axes. The resultant restoring forces are nonlinear with respect to the displacements in two axes. While the proposed novel 2d-NTMDI has the capability of mitigating bi-directional vibrations simultaneously, the system nonlinearity complicates the optimization. A numerical search method is adopted to determine the optimum design parameters of the 2d-NTMDI. To evaluate the effectiveness, the proposed 2d-NTMDI is deployed in a 5MW-wind turbine blade suffering from bi-directional vibrations in the edgewise and flapwise directions. To this end, an analytical model of a wind turbine blade with a 2d-NTMDI is established. The wind loading is computed using blade element momentum theory. The effectiveness of the 2d-NTMDI is examined under different loading conditions. Also, the fatigue damage of the wind turbine blades is calculated according to the rain-flow cycle counting approach and Miner's law. It is concluded that the optimized design 2d-NTMDI can prolong the blade fatigue life by $35\%$.  

\end{abstract}
\begin{keyword}
Two-dimensional response reduction, wind turbine blades, two-dimensional nonlinear tuned mass damper, fatigue damage mitigation
\end{keyword}

\end{frontmatter}
\section{Introduction}

The wind power industry is moving towards larger rotors to capture more energy and reduce the cost. The size increment of the wind turbine structure renders the blades more flexible and more susceptible to fatigue damage as a result of larger vibrations. Due to harsh aerodynamic loading, wind turbine blades always suffer from excessive vibrations which negatively influence the wind turbine performance and cause fatigue damage. To protect wind turbines, structural vibration control which has been effectively used in traditional civil structures \cite{Housner1997} is recently being studied to mitigate the vibrations of wind turbines.

Extensive studies have been carried out on reducing vibrations of wind turbine tower, or responses of floating wind turbine platforms. Murtagh \textit{et al.} \cite{Murtagh2007} deployed a singular tuned mass damper (TMD) to mitigate the vibrations of a wind turbine. The authors found that the TMD is effective when it is tuned to the dominant frequency of the wind turbine. Laface \textit{et al.} \cite{laface} compared the performance of a novel 2 degree of freedom (DOF) extended linear TMD with a 1-DOF linear TMD in reducing the motion of a spar type floating wind turbine (FWT). The authors concluded that the proposed 2-DOF TMD outperforms the traditional 1-DOF TMD in suppressing the motions of the FWT platform. Multiple linear TMDs have also been used in OWTs for increasing the robustness and improving the vibration mitigation of single TMDs \cite{Chen}. Xie \textit{et al.} \cite{Xie} compared the effectiveness of multiple linear TMDs with singular linear TMD in reducing the vibrations of a monopile fixed-bottom OWT in the presence of soil-structure interactions and concluded that the robustness of multiple TMDs is higher than single TMD. In addition to linear TMDs, tuned liquid dampers are also used to mitigate the vibrations of wind turbines. Ghaemmaghami \textit{et al.} \cite{Ghaemmaghami2013} used a tuned liquid damper in a numerical model of a wind turbine, and the authors concluded that the tuned liquid damper can effectively reduce the responses of wind turbine tower. Recently, tuned mass damper inerters (TMDI) have been proposed to reduce the vibrations of different structures with smaller mass in comparison with regular TMD. In TMDIs, the linear motion is converted to rotational motion which can considerably increase the mass of the controller \cite{Cao2019}. Sarkar and Fitzgerald \cite{Sarkar2019} utilized TMDI to mitigate the responses of the tower of a spar type FWT. It was observed that the TMDI can provide comparable response mitigation as regular TMDs and reduce the stroke. 

Although passive controlling techniques are promising in mitigating structural responses, they might lose part of their effectiveness in the presence of damages or uncertainties. To overcome this limitation, numerous studies have proposed using active and semi-active controllers \cite{Satish1, Satish2, Gupta, Dogruer}. Sun \cite{Sun2018,Sun2018_2} proposed using a semi-active TMD to mitigate the vibrations of the OWTs subjected to multi-hazard loading conditions. The author found that the semi-active TMD is effective in reducing the vibrations of the tower in fore-aft direction in the presence of soil effects and damages in the foundation and the tower. Sarkar and Chakraborty \cite{Sarkar2019_2} proposed using magneto-rheological fluid inside tuned liquid column damper (TLCD) to mitigate the wind turbine tower's vibrations. The authors concluded that using magneto-rheological fluid inside a TLCD can improve the effectiveness of the TLCD in mitigating the wind turbine vibrations. Fath \textit{et al.} \cite{Fath} deployed a tuned liquid multi-column damper integrated with three flow control valves for allowing the liquid to transfer between columns in a semi-submersible FWT. The authors demonstrated that increasing the mass ratio and cross-section of the tuned liquid multi-column damper improves the performance of the controller in stabilizing the FWT.

While the aforementioned studies focus mainly on reducing uni-directional vibrations (in-plane or out-of-plane), OWTs suffer from multi-directional vibrations due to vortex induced vibrations or misaligned wind and wave loadings. In this regard, Tong \textit{et al.} \cite{Tong2018} proposed using a bi-directional TLCD to reduce the motion of a barge type FWT in pitch and roll directions. The authors concluded that the bi-directional TLCD effectively reduces the motion of the barge in pitch and roll directions and improves the generated electrical power quality. Sun and Jahangiri \cite{Vahid:2018:MSSP} proposed a three-dimensional pendulum tuned mass damper (3d-PTMD) to reduce the bi-directional vibrations of the fixed-bottom OWT. The authors concluded that the 3d-PTMD can effectively reduce the tower responses. Furthermore, the proposed 3d-PTMD was used to reduce fatigue damage of OWTs \cite{Vahid:2018:ES}, and it was found that the 3d-PTMD can reduce the fatigue damage of the tower effectively. Jahangiri and Sun \cite{Vahid:2019:OE} proposed using an electromagnetic energy harvester instead of viscous damper to generate electrical power from the kinetic energy of the pendulum. It was found that, electrical energy in the magnitude of KWs can be captured using the electromagnetic energy harvester. To reduce the pendulum damper stroke and to increase the robustness of the 3d-PTMD, Jahangiri and Sun \cite{PPTMD} proposed using a three-dimensional pounding pendulum TMD in top of the wind turbine to reduce the bi-directional vibrations of the wind turbine tower. The 3d-PTMD is a passive controller and it might lose its effectiveness in the presence of system and environmental variations. Therefore, Sun \textit{et al.} \cite{Vahid:2021:SCHM} proposed using a three-dimensional adaptive pendulum TMD (3d-APTMD) to control the bi-directional responses of the tower under time-varying structural properties. The authors compared the effectiveness of the 3d-APTMD with 3d-PTMD and concluded that the 3d-APTMD is more effective in reducing the tower vibrations in the presence of structural property variations. To mitigate the three-dimensional vibration of a spar floating offshore wind turbine, Jahangiri and Sun  ~\cite{Vahid:2020:OE} proposed using a 3d-PTMD inside the nacelle to reduce the tower bi-directional vibrations and a dual linear pounding TMD in the floater to reduce the responses of the floater in roll and pitch directions. The dual linear pounding TMD stroke was limited to the available space inside the platform and was found to be effective in reducing the responses of the floater in roll and pitch directions.

Vibration mitigation of wind turbine towers and floating platforms have received considerable research efforts, yet studies on wind turbine blade vibration control are limited. As a critical component of wind turbines, blades suffer from vibrations in two directions: edgewise and flapwise. Zhang \textit{et al.} \cite{Zhang2015} proposed using TLCD to reduce the blade vibrations in the edgewise direction. It was concluded that the effectiveness of the controller is increased by increasing the mass ratio, and employing the controller near the blade tip can increase the damper effectiveness. Zhang and Fitzgerald \cite{Zhang2020} utilized a TMDI to mitigate the vibrations of the blades in the edgewise direction. It was found that the TMDI requires a much smaller stroke than classical linear TMDs. A few studies have focused on reducing the vibrations of the blade in flapwise direction considering the limited space inside the blade in the out-of-plane direction. Arrigan \textit{et al.} \cite{Arrigan2011} proposed semi-active TMDs and concluded that semi-active TMDs can effectively mitigate the responses of wind turbine blades in flapwise direction under varying operational and environmental conditions. Zuo \textit{et al} \cite{Zuo2017} used multiple TMDs to mitigate the vibrations of the blade in flapwise direction. It was shown that multiple TMDs can effectively reduce the flapwise vibrations. However, it was found that a single TMD is more effective than multiple TMDs.

To reduce multi-modal vibrations, Viet and Nghi \cite{Viet} proposed using a nonlinear single mass two-frequency pendulum TMD to reduce the response in the horizontal direction. The proposed device was shown effective in reducing two-mode vibrations of the primary structure. Almzan \textit{et a.} \cite{Almazan} proposed using a bi-directional and homogeneous TMD to reduce bi-directional vibrations with the capability of tuning the controller in each direction independently. Recently, Ohsaki \textit{et al.} \cite{Ohsaki} proposed using a tetrahedral tuned mass damper to reduce three-directional seismic responses. The authors optimized the parameters of the controller and demonstrated the effectiveness of the controller in reducing three-directional vibrations. In a recent study, Jahangiri and Sun \cite{Vahid_3dTMD} developed a novel three dimensional nonlinear TMD (3d-NTMD) with the capability of reducing three-dimensional structural responses. The authors examined the effectiveness of the 3d-NTMD in reducing the motion of FWTs in pitch, roll and heave directions at the same time under parked and operating condition. It was shown that the 3d-NTMD is effective in suppressing the three-dimensional responses of the FWT. Wind turbine blades vibrate in both directions of flapwise and edgewise simultaneously. However, to the author's knowledge, none of the existing references have addressed the bi-directional vibration mitigation of wind turbine blades.

In this paper, a novel two-dimensional nonlinear tuned mass damper inerter (2d-NTMDI) is proposed to reduce the bi-dimensional vibrations of a primary structure. The 2d-NTMDI consists of a single mass and two pairs of dash-pots, springs and inerters. The resultant restoring force of the 2d-NTMDI is characterized by significant geometric nonlinearity because of the springs' configurations. The proposed 2d-NTMDI has potential benefits over existing TMDs in two aspects. First, it can reduce  the responses of the primary structure in two directions at the same time. Second, the 2d-NTMDI has the capability of reducing multi-mode vibrations and it can be tuned to the dominant frequency of the primary structure in each direction independently. A numerical search method is adopted to determine the optimum design parameters of the 2d-NTMDI. To evaluate the effectiveness of the proposed 2d-NTMDI, the 5MW wind turbine blades are used and subjected to excessive vibrations in two directions: edgewise and flapwise. To this end, an analytical model of a wind turbine blade with a 2d-NTMDI is established. The wind loading is computed using blade element momentum theory. The effectiveness of the 2d-NTMDI is examined under different loading conditions. Also, the fatigue damage of the wind turbine blades is calculated according to the rain-flow cycle counting approach and Miner's law. It is found that an optimized 2d-NTMDI can prolong the blade fatigue life by around $35\%$.
\section{Mathematical model of 2d-NTMDI}\label{model}

\begin{figure}[ht]
\centering
 \includegraphics[scale=0.3]{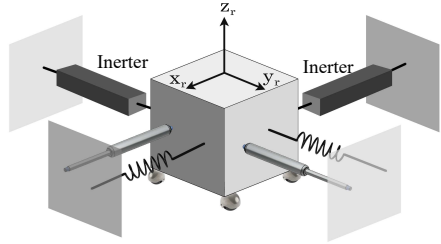}
 \caption{Schematic model of the 2d-NTMDI}
\label{fig 2d-NTMDI}
\end{figure}

A schematic model of the 2d-NTMDI is demonstrated in Fig. \ref{fig 2d-NTMDI}. The developed 2d-NTMDI consists of a mass with two sets of springs, dash-pots and inerters in two axes $X$ and $Y$. Due to geometric nonlinearity of the spring configuration, resultant restoring force of the damper is nonlinear with respect to the displacements in the two directions, thereby complicating the optimal design of the 2d-NTMDI.

\begin{figure}[ht]
	\centering
  \includegraphics[scale=0.45]{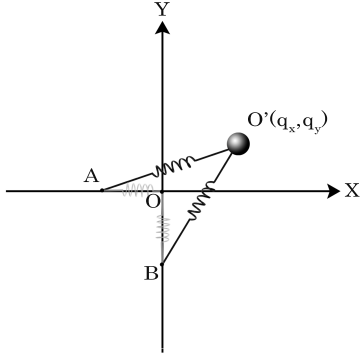}
	\caption{Configuration of 2d-NTMD at an arbitrary location}
	\label{2d-NTMDI_point}
\end{figure}

The non-linear spring forces can be determined by considering the movement of the TMD mass block from node O to O’ as illustrated in Fig. \ref{2d-NTMDI_point}. Deformation of each spring can be written as follows:
\begin{eqnarray}
\delta_x = O'A-OA = \sqrt{(l+q_x)^2+q_y^2}-l  \nonumber \\
\delta_y = O'B-OB = \sqrt{q_x^2+(l+q_y)^2}-l \label{eq10_1}  
\end{eqnarray}
where $l$ is the original length of each spring; $q_x$, and $q_y$ represent the relative coordinates of the TMD mass in $X$ and $Y$ directions with respect to the primary structure. 

The restoring force vector provided by each spring can be determined as:
\begin{eqnarray}
\overrightarrow{f_{k_x}} = k_x\delta_x\frac{\overrightarrow{r_{AO'}}}{|r_{AO'}|} \label{eq10_2} \\ 
\overrightarrow{f_{k_y}} = k_y\delta_y\frac{\overrightarrow{r_{BO'}}}{|r_{BO'}|}   \nonumber 
\end{eqnarray}
where $k_x$ and $k_y$ denote the stiffness coefficients of each spring in $X$ and $Y$ directions. $\overrightarrow{r_{AO'}}$ and $\overrightarrow{r_{BO'}}$ are the displacement vectors from point $O'$ to points $A$ and $B$ respectively, which can be expressed as:
\begin{eqnarray}
\overrightarrow{r_{AO'}} = (l+q_x)\vec{i} + q_y\vec{j} \label{eq10_3}  \\ 
\overrightarrow{r_{BO'}} = q_x\vec{i} + (l+q_y)\vec{j}    \nonumber
\end{eqnarray}

By substituting Eqs.(\ref{eq10_1}) and (\ref{eq10_3}) into Eq.(\ref{eq10_2}), the restoring force can be obtained as:
\begin{eqnarray}
f_{k_x}(q_x,q_y) &=& k_x(l+q_x)\left(1-\frac{l}{\sqrt{(l+q_x)^2+q_y^2}}\right) \nonumber \\ 
&+& k_y(q_x)\left(1-\frac{l}{\sqrt{q_x^2+(l+q_y)^2}}\right) \label{eq10_4_a}  \nonumber \\  \nonumber \\
f_{k_y}(q_x,q_y) &=& k_x(q_y)\left(1-\frac{l}{\sqrt{(l+q_x)^2+q_y^2}}\right) \\ \nonumber
&+& k_y(l+q_y)\left(1-\frac{l}{\sqrt{q_x^2+(l+q_y)^2}}\right)   \label{eq10_4_b}
\end{eqnarray}

\begin{figure}[ht]
	\centering
  \includegraphics[scale=0.25]{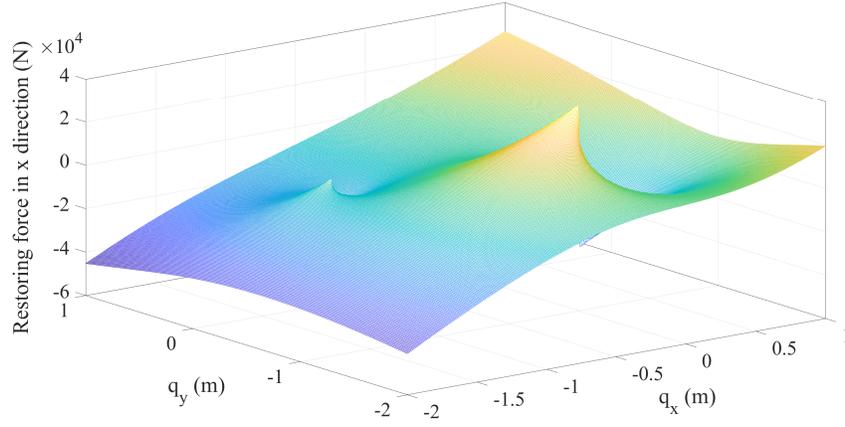}
	\caption{Restoring force-displacement surface plot of the 2d-NTMDI}
	\label{fig_force}
\end{figure}

As a representative case, the force-displacement surface of $f_{k_x}(q_x,q_y)$ is shown in Fig. \ref{fig_force}, where the system exhibits significant nonlinearity when $q_x$ and $q_y$ are within a certain range.

\begin{figure}[ht]
	\centering
  \includegraphics[scale=0.25]{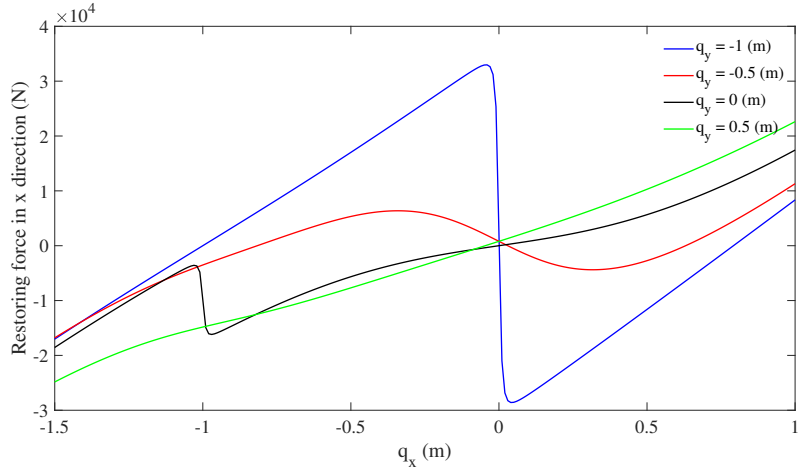}
	\caption{Restoring force-displacement curve under fixed relative displacement in $y$ direction($q_y = -1, -0.5, 0, 0.5$)}
	\label{fig_force_X}
\end{figure}

Similarly, the force-displacement curve of $f_{k_x}(q_x,q_y)$ against relative displacement in $X$ direction is shown in Fig. \ref{fig_force_X} where $q_y = -1, -0.5, 0, 0.5$. The nonlinear restoring force of the system can be further observed in Fig. \ref{fig_force_X}.

\begin{figure}[ht]
	\centering
  \includegraphics[scale=0.55]{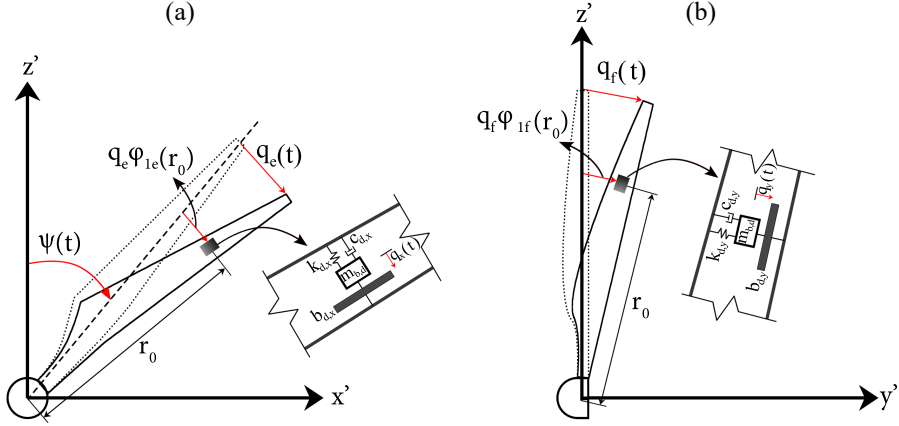}
	\caption{Coordinates of the turbine blades in (a) edgewise and (b) flapwise directions with the 2d-NTMDI}
	\label{Blade_2d-NTMDI}
\end{figure}

To evaluate the effectiveness of the proposed 2d-NTMDI in reducing the vibrations of the wind turbine blade in edgewise and flapwise, a 5 MW baseline wind turbine blade is adopted and the 2d-NTMDI is deployed in the blade, as shown in Fig. \ref{Blade_2d-NTMDI}. The schematic model of a reduced 4 degree of freedom wind turbine blade coupled with the 2d-NTMDI is illustrated in Fig. \ref{Blade_2d-NTMDI}. The edgewise coordinate of the blade is shown with $q_e$ and the flapwise coordinate is represented by $q_f$. Also, the relative coordinates of the 2d-NTMDI with respect to the blade is shown by $q_x$ and $q_y$.

The position of the 2d-NTMDI can be written as:
\begin{eqnarray}
x'_{r,d} &=& r_0\sin\psi_1 + q_x\cos\psi_1 + q_e\phi_{1e}(r_0)\cos \psi_1 \nonumber \\
y'_{r,d} &=& q_y + q_f\phi_{1f}(r_0)\\
z'_{r,d} &=& r_0\cos\psi_1 - q_x\sin\psi_1 - q_e\phi_{1e}(r_0)\sin \psi_1\nonumber
\label{eq10_5}
\end{eqnarray}
where variable $r_0$ is the location of the 2d-NTMDI from the root of the blade. $\psi_1$ is the azimuthal angle of the $1^{st}$ blade. $\phi_{1e}$ and $\phi_{1f}$ represent the first mode shape of the blade in edgewise and flapwise directions respectively. 

Taking the first derivative of the 2d-NTMDI coordinates gives the velocity components as: 
\begin{eqnarray}
\dot{x'}_{r,d} &=& \Omega r_0\cos\psi_1 + \dot{q_e}\phi_{1e}(r_0)\cos\psi_1 - \Omega q_e\phi_{1e}(r_0)\sin\psi_1  \nonumber \\
 &+& \dot{q}_x\cos\psi_1 - \Omega q_x \sin \psi_1 \nonumber \\
\dot{y'}_{r,d} &=& \dot{q_y} + \dot{q_f}\phi_{1f}(r_0) \\
\dot{z'}_{r,d} &=& -\Omega r_0\sin \psi_1 - \dot{q_e}\phi_{1e}(r_0)\sin\psi_1 - \Omega q_e\phi_{1e}(r_0)\cos\psi_1 \nonumber \\
&-& \dot{q}_x\sin \psi_1 - \Omega q_x \cos \psi_1 \nonumber
\label{eq10_6}
\end{eqnarray}

The absolute velocity magnitude of the 2d-NTMDI is: 
\begin{equation}
v_{b,d}(r,t) = \sqrt{\dot{x'}_{r,d}^2 + \dot{y'}_{r,d}^2 + \dot{z'}_{r,d}^2}
\label{eq10_7}
\end{equation}

The kinetic energy of the 2d-NTMDI can be expressed as follows:
\begin{equation}
T_{b,d} = \frac{1}{2}m_{b,d}v_{b,d}^2 + \frac{1}{2}b_{d,x} \dot{q_x^2} + \frac{1}{2}b_{d,y} \dot{q_y^2}
\label{eq10_8}
\end{equation}
where $m_{b,d}$ is the mass of the 2d-NTMDI; $b_{d,x}$ and $b_{d,y}$ are the inertance in $X$ and $Y$ directions. The total kinetic energy of the 4-DOF wind turbine blade 2d-NTMDI system can be written as:

\begin{equation}
T = \frac{1}{2}\int_{0}^{R}{\bar{m}v^2_{b1}(r,t)dr} + T_{b,d}
\label{eq10_9}
\end{equation}
where $v^2_{b1}$ is the absolute velocity magnitude of the unit $dr$ for the first blade. More details on the absolute velocity  $v_{b1}$ can be found in \cite{Vahid:2018:MSSP}. The potential energy of the 2d-NTMDI can be determined as follows:
\begin{equation}
V_{b,d} = m_{b,d} g (r_0 \cos \psi_1 - \phi_{1e}(r_0) q_x \sin \psi_1 - q_e\sin \psi_1) + V_{restore}
\label{eq10_10}
\end{equation}
where $V_{restore}$ is the potential energy of the 2d-NTMDI caused by the nonlinear springs. The total potential energy of the 4-DOF wind turbine blade system can be calculated by adding $V_{b,d}$ to the potential energy of the blade, which can be found in more details in \cite{Vahid:2018:MSSP}.

By substituting the kinetic energy and potential energy equations into the Euler-Lagrangian equation, the system’s equations of motion can be derived and written in a matrix format as follows:

\begin{equation}
\tilde{M}\ddot{\tilde{q}} + \tilde{C}\dot{\tilde{q}} + (\tilde{K})\tilde{q} = \tilde{Q}_{wind} + \tilde{Q}_{V,b}
\label{eq10_11}
\end{equation}
where $\tilde{M}$, $\tilde{C}$  and $\tilde{K}$ are the 4-DOF wind turbine blade system's mass matrix, damping matrix and stiffness matrix, which have a dimension of $4 \times 4$ which are expressed as follow:

\begin{equation}
\tilde{M} = \left[
\begin{array}{cccc}
m_{1} + a^2 m_{b,d}  & 0 & a m_{b,d} & 0\\
0 & m_{2} + b^2 m_{b,d} & 0 & b m_{b,d}\\
a m_{b,d} & 0 &  m_{b,d}+b_{d,x} & 0\\
0 & b m_{b,d} & 0 &  m_{b,d}+b_{d,y}\\
\end{array}
\right]
\end{equation}

\begin{equation}
\tilde{K} = \left[
\begin{array}{cccc}
k_{b,eg} - a^2\Omega^2m_{b,d}  & 0 & -a \Omega^2 m_{b,d} & 0\\
0 & k_{b,fp} & 0 & 0\\
-a \Omega^2 m_{b,d} & 0 &  k_{d,x} - \Omega^2 m_{b,d} & 0\\
0 & 0 & 0 &  k_{d,y}\\
\end{array}
\right]
\end{equation}

\begin{equation}
\tilde{C} = \left[
\begin{array}{cccc}
c_{b,eg}  & 0 & 0 & 0\\
0 & c_{b,fp} & 0 & 0\\
0 & 0 &  c_{d,x} & 0\\
0 & 0 & 0 &  c_{d,y}\\
\end{array}
\right]
\end{equation}
where $a = \phi_{1e}(r_0)$, $b = \phi_{1f}(r_0)$, $m_{1} = \int_0^R\bar{m}(r)\phi^2_{1e}dr$, and $m_{2} = \int_0^R\bar{m}(r)\phi^2_{1f}dr$. Also, $k_{b,eg} = k_{eg} + k_{ge,eg} - k_{gr,eg}cos\psi -\Omega^2 m_1$ and  $k_{b,fp} = k_{fp} + k_{ge,fp} - k_{gr,fp}cos\psi$. Variable $\tilde{Q}_{wind}$ is the generalized force vectors corresponding to wind loading. $\tilde{Q}_{V,b}$ is derived from the potential energy of the 2d-NTMDI and is equal to:
\begin{equation}
\tilde{Q}_{V,b} = \left\{
\begin{array}{cccc}
m_{b,d}ga\sin\psi & 0 & m_{b,d}g\sin\psi & 0
\end{array}
\right\}^T
\end{equation}

\section{Aerodynamic loading}

\begin{figure}[ht]
	\centering
  \includegraphics[scale=0.23]{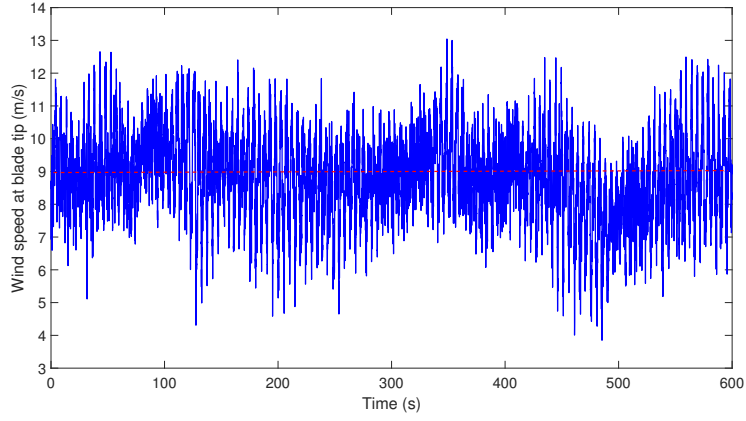}
	\caption{Wind speed at tip element of wind turbine blade}
	\label{fig wind speed}
\end{figure}

This section presents the derivation of the generalized wind loading, which is denoted by $\tilde{Q}_{wind}$ in Eq. (\ref{eq10_11}). It is noted that a three-dimensional wind field profile in this study is computed using the IEC Kaimal spectral model and the logarithmic wind field profile. Then the wind speeds at different locations on the blades are calculated. Detailed information with respect to the generation of the three-dimensional wind field and calculation of wind speeds on the blades can be found in the authors' earlier work \cite{Sun2018, Vahid:2018:MSSP, Vahid:2020:OE}. To save space, a concise description of the aerodynamic loading is presented in this section. As a representative case, the wind speed history at the tip element of wind turbine blade one is shown in Fig. (\ref{fig wind speed}) where the mean wind speed is $9$ $m/s$ and the  turbulent intensity is $15\%$. 

\begin{figure}[ht]
	\centering
  \includegraphics[scale=0.25]{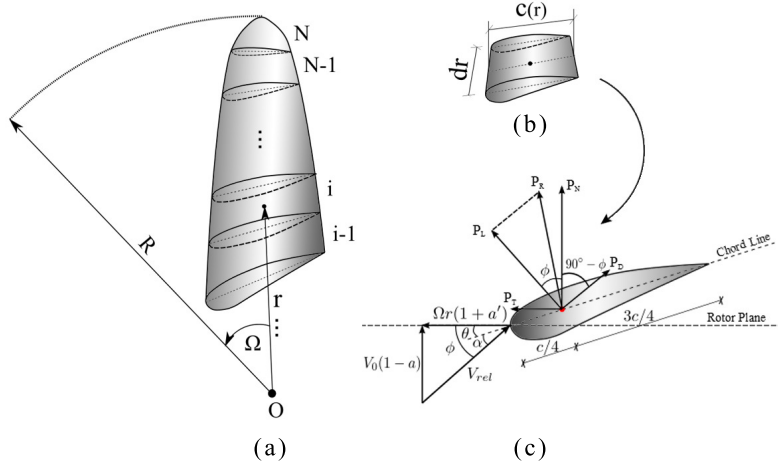}
	\caption{(a): Wind turbine blade discretized into $N$ blade elements. (b); $i^{th}$ blade element. (c): Blade element section}
	\label{fig blade}
\end{figure}

The wind loading applied to the wind turbine blade is calculated using the Blade Element Momentum (BEM) method. As shown in Fig. \ref{fig blade} (a), the wind turbine blade is discretized into $N$ elements. Details of the $ith$ blade element, which is placed at a distance $r$ from the blade root, is shown in Fig. \ref{fig blade} (b), where $dr$ denotes the element span and $c(r)$ is the chord length at the element mid-span.

The normal and tangential forces acting on each segment of the blade can be computed as:
\begin{equation}
P_N = \frac{1}{2}\rho V^2_{rel}c C_N,\quad P_T = \frac{1}{2}\rho V^2_{rel}c C_T  
\label{eq16}
\end{equation}
where $C_N$ and $C_T$ are the normal and tangential coefficients; Parameter $V_{rel}$ denotes the relative wind speed which can be computed as follows:
\begin{equation}
V_{rel} = \sqrt{[v(1-a)]^2 + [\Omega r (1+a')]^2} 
\label{eq17}
\end{equation}
where parameters $a$ and $a'$ represent the axial speed and tangential speed induction factors.

Based on the principles of work and energy, the generalized aerodynamic loading can be defined as:

\begin{eqnarray}
Q_{e} &=& \int_0^R P_{T}(r,t)\phi_{1e}dr, \: Q_{f} = \int_0^R P_{N}(r,t)\phi_{1f}dr
\label{eq18}
\end{eqnarray} 

\begin{figure}[ht]
	\centering
  \includegraphics[scale=0.30]{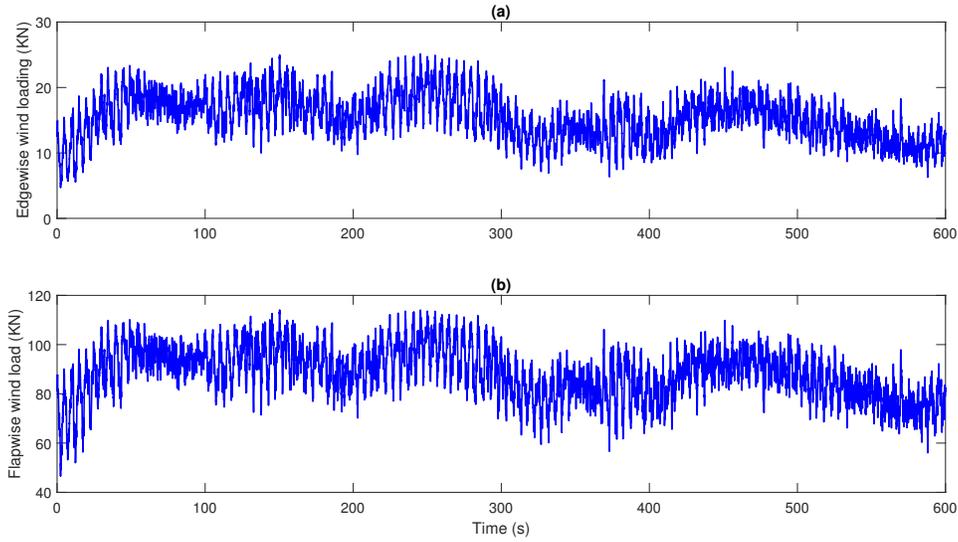}
	\caption{Generalized wind loading acting on wind turbine blade in (a): Edgewise direction. (b): Flapwise direction with a $9$ $m/s$ mean wind speed an $15\%$ turbulent intensity.}
	\label{fig wind load}
\end{figure}

The edgewise and flapwise loading acting on the wind turbine blade where the mean wind speed is assumed to be $9$ $m/s$ with a $15\%$ turbulent intensity is demonstrated in Fig. \ref{fig wind load} (a) and (b).

\section{Wind data}

According to \cite{Skipjack}, a new project with $15$ monopile fixed-bottom offshore wind turbines is being planned to be constructed in the upcoming years. To evaluate the performance of the 2d-NTMDI in reducing the fatigue damage of the wind turbine blades, real wind data of the last 25 years is considered.

\begin{figure}[ht]
	\centering
  \includegraphics[scale=0.23]{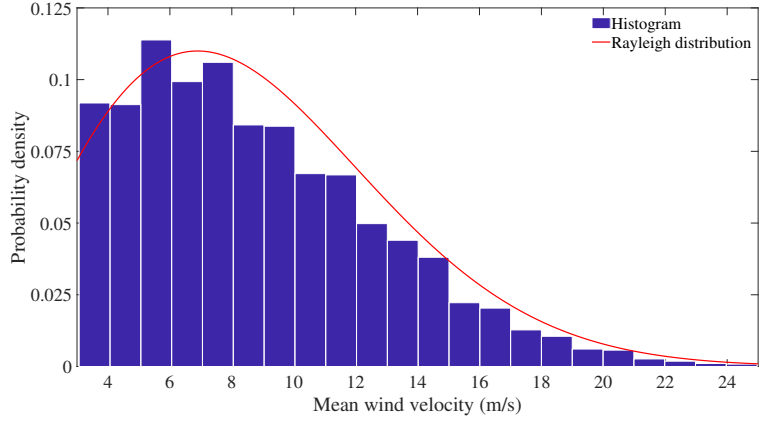}
	\caption{Wind speed histogram of station 44009 represented by Rayleigh distribution}
	\label{fig_wind}
\end{figure}

The histogram plot of wind velocity distribution of station 44009 is shown in Fig. \ref{fig_wind}. One can find that the wind velocity in the range of $[5-9]$ $m/s$, has a $55\%$ probability of occurrence. Also, it can be seen that wind speeds beyond $18$ $m/s$ have a relatively small probability of occurrence (around $3\%$). Based on the original wind speed data, a Rayleigh distribution can be used to approximate the wind velocity distribution as expressed as follows:

 \begin{equation}
f(x,\sigma) = \frac{x}{\sigma^2}e^{\frac{-x^2}{2\sigma^2}}
\label{eq10_12_rayleigh}
\end{equation}
where $x=6.81$ and $\sigma = 6.927$.

The wind data is divided into five mean wind velocity bins. A representative wind speed for each bin is chosen and the wind turbine blades are simulated under the representative wind speeds. For each chosen mean wind speed, a turbulence intensity is chosen. For $\bar{V} = 3 m/s$ a $TI=20\%$,  $\bar{V} = 6 m/s$ a $TI=20\%$, $\bar{V} = 9 m/s$ a $TI=15\%$, $\bar{V} = 12 m/s$ a $TI=10\%$ and $\bar{V} = 15 m/s$ a $TI=10\%$ is chosen.

\section{2d-NTMDI optimization}

A numerical search approach is carried out to determine the optimum parameters of the 2d-NTMDI. Reducing the fatigue damage in wind turbine blades is one of the most important concerns. Therefore, minimizing the combined root mean square (RMS) response of the blade in edgewise and flapwise directions is considered as the objective optimization function which is defined as:

\begin{equation}
J_{opt} = \sqrt{\alpha X_{RMS}^2 + \beta Y_{RMS}^2}
\label{eq10_12}
\end{equation}
where $X_{RMS}$ and $Y_{RMS}$ are the RMS response in the edgewise and flapwise directions respectively. $\alpha$ and $\beta$ are the weighting factors for the RMS response in edgewise and flapwise directions. Generally, the response in flapwise direction is larger than the response in edgewise direction due to the thrust force of the aerodynamic loading. Thus, a larger weighting factor is considered for RMS response in edgewise direction. In this research $\alpha=0.95$ and $\beta=0.05$ are adopted. It is noted that different weighting factors can be adopted in real applications to achieve different desired optimum objectives.

The parameters of the 2d-NTMDI which need to be optimized include mass ratio$(\mu)$, inertance ratios in $x$ direction $(\beta_x)$, and $y$ direction $(\beta_y)$, damping ratios in $x$ direction $(\zeta_x)$, and $y$ direction $(\zeta_y)$, frequency ratios in $x$ direction $(f_x)$, and $y$ direction $(f_y)$. The mass ratio$(\mu)$, inertance ratio$(\beta)$, and damping ratio$(\zeta)$ can be determined as:

\begin{equation}
\mu = \frac{m_{b,d}}{m_{b,0}}; \qquad \beta = \frac{b_d}{m_{b,0}}; \qquad \zeta = \frac{c_d}{2\sqrt{k_{b,d}(m_{b,d}+b_d)}}
\label{eq10_13}
\end{equation}
where $m_{b,0}$ is the blade first modal mass. $m_{b,d}$ represents the mass of the 2d-NTMDI  and $b_d$ denotes the actual mass of the inerter system in $kg$.

To reduce the computational cost caused by optimizing seven parameters ($\mu$, $\beta_x$, $\beta_y$, $\zeta_x$, $\zeta_y,$ $f_x$, and $f_y$), identical values of the damping ratio in $x$ and $y$ directions are used. Therefore, the number of design parameters is reduced to $5$, including mass ratio($\mu$), damping ratio($\zeta$), frequency ratio in x direction($f_x$) and $y$ direction ($f_y$).

Seven different mass ratios ranging from $2\% (28kg)$ to $8\% (112kg)$ are considered. For each mass ratio, the inertance ratio is chosen such that the damper stroke is within the available space inside the blade. It is important to mention that the ratio of the inerter's physical mass to the inertance coefficient ($b_d$) is around $\frac{1}{200}$ \cite{Inerter_mass}, which means that an inerter with $100 \%$ inertance ratio would have a physical mass of only $7 kg$, which is a relatively small value in comparison to the TMD mass.

The available space (chord length and thickness) at different locations inside the blade is tabulated in Table \ref{Tab_Blade_Space}. Preliminary simulations were carried out for the 2d-NTMDI placed at different locations, and the damper location is selected to be at $r_0=45$ $(m)$, where the blade chord length is $3.0$ $(m)$ and the thickness is $0.55$ $(m)$. The effectiveness of the 2d-NTMDI increases by moving closer to the tip of the blade. However, the available space for the 2d-NTMDI reduces which constraints the 2d-NTMDI and negatively affects the performance of the 2d-NTMDI. Therefore, $r_0=45$ $(m)$ is chosen as the location of the damper.
It is important to mention that the external loading affects the optimized design parameters of the 2d-NTMDI. By increasing the wind speed the damper stroke increases. Hence, it is important to optimize the design parameters to assure that the 2d-NTMDI operates within the available space inside the wind turbine blade. Also, considering the wind speed distribution presented in the previous section, the optimization is carried out under two representative wind velocities: $6$ $m/s$, and $9$ $m/s$ which represents around $55\%$ of the wind velocity statistics in the targeted marine area \cite{Vahid:2018:ES}.

\begin{center}
\begin{table*}[t]%
\centering
\caption{Available space inside the NREL 5MW wind turbine blade \cite{Sandia}\label{Tab_Blade_Space}}%
\begin{tabular*}{\columnwidth}{@{\extracolsep\fill}lccc@{\extracolsep\fill}}%
\toprule
\textbf{$r_0 (m)$} & \textbf{Chord length (m)}  & \textbf{Thickness (m)}  \\
\midrule
20 & 4.5  & 1.6   \\
30 & 3.9  & 1.0    \\
45 & 3.0  & 0.55 \\
55 & 2.3  & 0.42 \\
61 & 1.4 & 0.25 \\
\bottomrule
\end{tabular*}
\end{table*}
\end{center}

\subsection{Optimization under $\bar{V}$ = 6 $m/s$}

\begin{figure}[ht]
	\centering
  \includegraphics[scale=0.26]{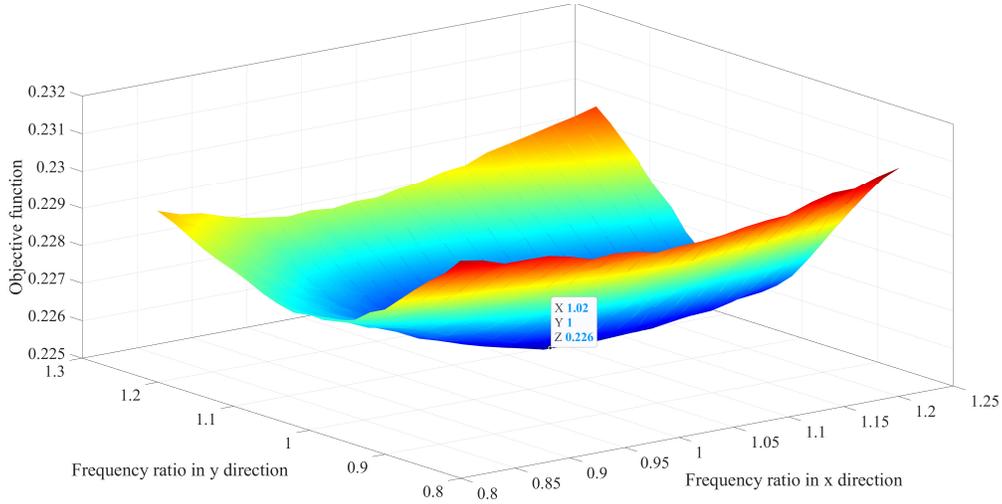}
	\caption{Objective function surface under different frequency ratios for $\mu = 4\%$, $\zeta = 16\%$, $\beta_x = 5\%$, and $\beta_y = 80\%$ under a mean wind speed of $6$ $m/s$.}
	\label{2dNTMDI_OPT_V6_3D}
\end{figure}

In this subsection, the optimized parameters of the 2d-NTMDI will be determined for different mass ratios under $\bar{V} = 6 m/s$ and a $20\%$ turbulence intensity. Fig. \ref{2dNTMDI_OPT_V6_3D} illustrates the objective optimization function against different frequency ratios in $x$ and $y$ directions for a representative case of $\mu = 4\%$, $\zeta = 16\%$, $\beta_x = 5\%$, and $\beta_y = 80\%$. Based on the preliminary results, the desired frequency ratio lies in a range of $[0.8 \quad 1.2]$ and the damping ratio lies in a range of $[9\% \quad 20\%]$. The frequency ratio, damping ratio and inertance ratio values that produce the minimum value of the objective function within the available space inside the blade are chosen as the optimum design parameters.

\begin{figure}[ht]
	\centering
  \includegraphics[scale=0.25]{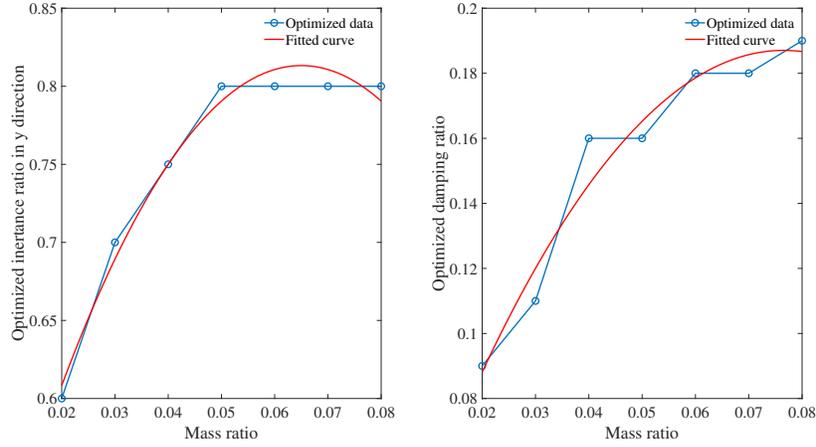}
	\caption{optimum damping ratio and inertance ratio in $y$ direction under a mean wind speed of $6$ $m/s$.}
	\label{2dNTMDI_V6_B_Z}
\end{figure}

Fig. \ref{2dNTMDI_V6_B_Z} illustrates the optimum damping ratio and inertance ratio values in $y$ direction under different mass ratios. It can be observed that the optimized damping ratio and inertance ratio increases with the increase of mass ratio. To facilitate engineering application, the optimum design formula for inertance ratio and damping ratios are derived via fitting the data points in Fig. \ref{2dNTMDI_V6_B_Z}. It is important to note that different inertance ratios in $x$ direction have been tested and it is found that a $5\%$ inertance ratio in $x$ direction can optimally provide mitigation to reduce the RMS response of the blade in edgewise direction for all the considered mass ratios.

\begin{eqnarray}
\zeta_{opt} &=& - 30.95\mu^2 + 4.7381\mu + 0.0057 \nonumber \\ \label{eq10_14}
\beta_{y,opt} &=& - 100.1\mu^2 + 13.1\mu + 0.39   
\end{eqnarray}

\begin{figure}[ht]
	\centering
  \includegraphics[scale=0.25]{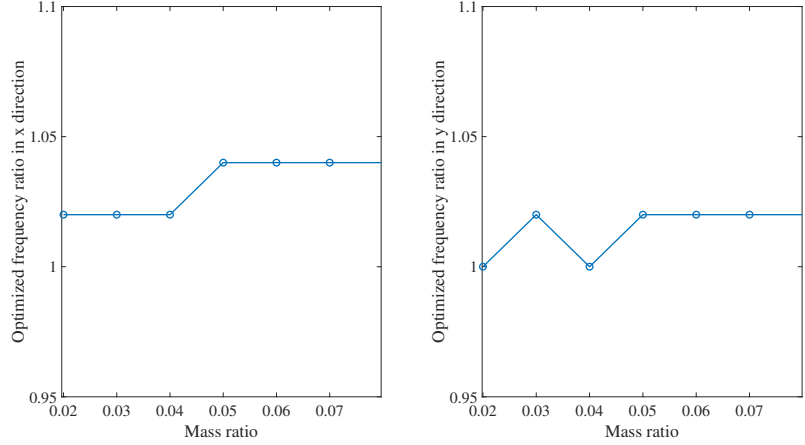}
	\caption{optimum design of frequency ratios in $x$ and $y$ directions under a mean wind speed of $6$ $m/s$.}
	\label{2dNTMDI_V6_f}
\end{figure}

Fig. \ref{2dNTMDI_V6_f} portrays the optimum frequency ratios in $x$ and $y$ directions under different mass ratios. It can be seen that the frequency ratio in $x$ and $y$ directions is in the range of $[1.00 - 1.04]$ for different mass ratios.

\subsection{Optimization $\bar{V}$ = 9 $m/s$}

\begin{figure}[ht]
	\centering
  \includegraphics[scale=0.26]{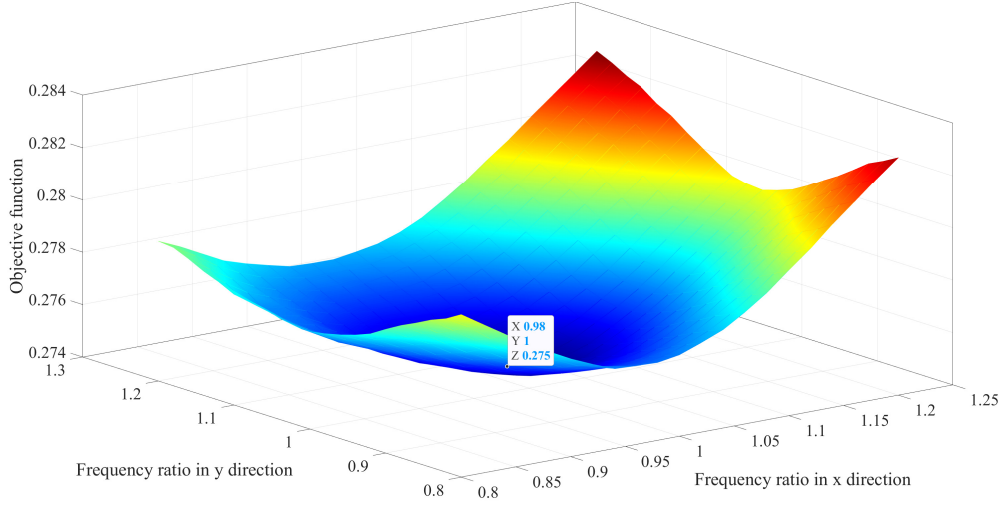}
	\caption{Objective function surface under different frequency ratios for $\mu = 4\%$, $\zeta = 15\%$, $\beta_x = 5\%$, and $\beta_y = 85\%$ under a mean wind speed of $9$ $m/s$.}
	\label{fig 2dNTMDI_OPT_V9_3D}
\end{figure}

Similarly, the optimized parameters of the 2d-NTMDI are determined for different mass ratios under $\bar{V} = 9$ $m/s$ and a $15\%$ turbulence intensity. Fig. \ref{fig 2dNTMDI_OPT_V9_3D} illustrates the objective optimization function against different frequency ratios in $x$ and $y$ directions for a representative case of $\mu = 4\%$, $\zeta = 15\%$, $\beta_x = 5\%$, and $\beta_y = 85\%$. Preliminary simulations have been carried out for different values of frequency and damping ratios and it is observed that the optimized frequency ratio lies in a range of $[0.8 \quad 1.2]$ and the optimized damping ratio lies in a range of $[9\% \quad 20\%]$. The frequency ratio, damping ratio and inertance ratio values that produce the minimum value of objective function within the available space inside the blade are chosen as the optimum design parameters.

\begin{figure}[ht]
	\centering
  \includegraphics[scale=0.3]{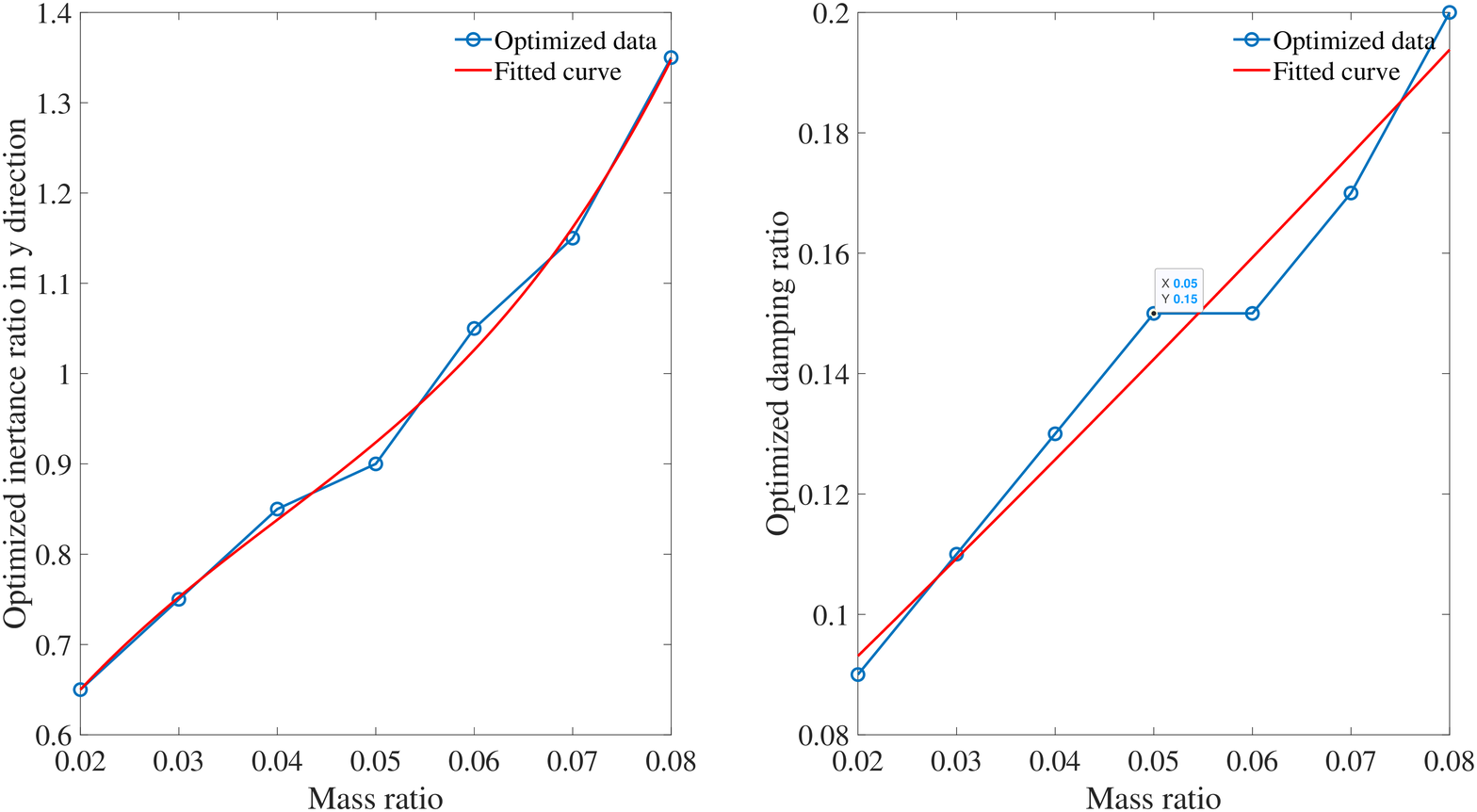}
	\caption{optimum design values for damping ratios and inertance ratio in $y$ direction under $9$ $m/s$ wind velocity}
	\label{fig 2dNTMDI_V9_B_Z}
\end{figure}

Fig. \ref{fig 2dNTMDI_V9_B_Z} illustrates the optimum damping ratios and inertance ratio in $y$ direction under different mass ratios. It can be observed that the optimized damping ratio and inertance ratio values increases with the increase of mass ratio. To facilitate engineering application, the optimum design formula for inertance ratio and damping ratios is derived via fitting the data points in Fig. \ref{fig 2dNTMDI_V9_B_Z}.

\begin{eqnarray}
\beta_{y,opt} &=& 2778\mu^3 - 333.3\mu^2 + 21.63\mu + 0.3286 \nonumber \\ \label{eq10_15}
\zeta_{opt} &=& 1.2\mu^2 + 1.6\mu + 0.061  
\end{eqnarray}

\begin{figure}[ht]
	\centering
  \includegraphics[scale=0.3]{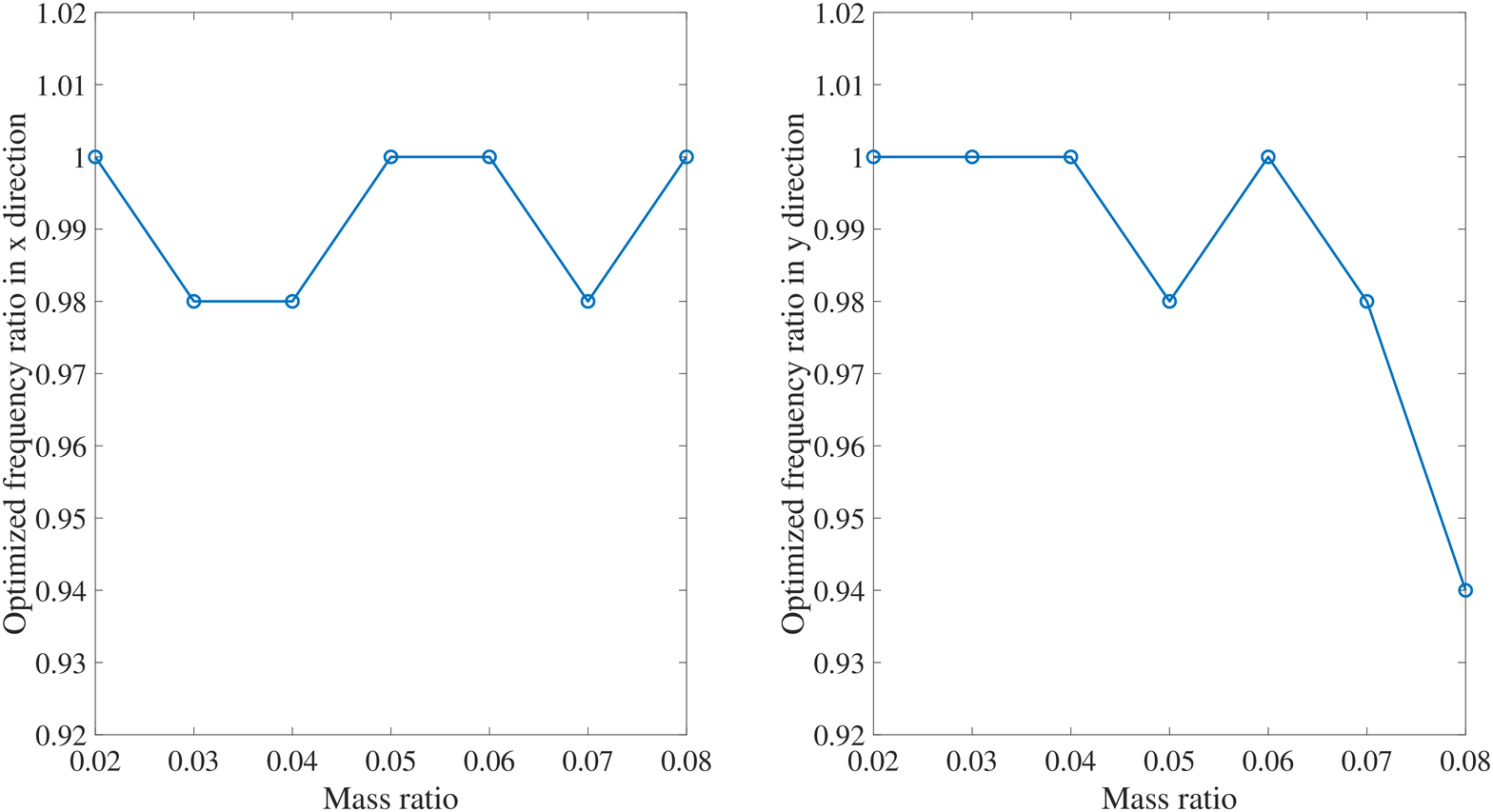}
	\caption{optimum design values for frequency ratios in $x$ and $y$ directions under $9$ $m/s$ wind velocity}
	\label{fig 2dNTMDI_V9_f}
\end{figure}

Fig. \ref{fig 2dNTMDI_V9_f} portrays the optimum frequency ratios in $x$ and $y$ directions under different mass ratios. It can be seen that the frequency ratio in $x$ and $y$ directions is in the range of $[0.94 - 1.00]$ for different mass ratios.

\section{Fatigue damage estimation}

The wind turbine blades' fatigue damage under the recorded wind data is estimated via the rain-flow cycle counting approach.

The number of cycles to failure caused by fatigue damage can be calculated as \cite{Sandia}:

 \begin{equation}
N_F = (\frac{S_i\gamma}{C})^{-b}
\label{eq10_12_1}
\end{equation}
where $\gamma$ is the safety factor which is equal to $1.35$ according to \cite{Sandia}. $S_i$ is the stress range caused by the bending moment. $C$ and $b$ are the material properties which can be found through the S-N curve of the considered material. The focus of this study is to estimate the fatigue damage at the blade root. According to \cite{Sandia}, SNL(Triax) material is used at the wind turbine blade root, whereas $C = 700$ $MPa$ and $b = 10$.

Total fatigue damage can be estimated using the miner's rule expressed as follows:

\begin{equation}
D = \Sigma_i {\frac{n_i}{N_F}}
\label{eq10_12_2}
\end{equation}
where $n_i$ is the number of cycles at stress range of $S_i$. 

Five different wind speeds of $3 m/s$, $6 m/s$, $9 m/s$, $12 m/s$, and $15 m/s$ are considered to estimate the wind turbine blade fatigue damage. The structural response of the blade - 2dNTMDI system is simulated for each loading scenario and the corresponding fatigue damage is estimated via the rain-flow cycle counting approach. Finally, the cumulative fatigue damage is estimated as:

\begin{equation}
D_{total} = \Sigma_{i} P_iD_i
\label{eq10_12_3}
\end{equation}
where probability of each loading condition is represented by $P_i$ and $D_i$ denotes the associated fatigue damage.

\section{Results and Discussions}

In this section, response mitigation of the wind turbine blade under two different representative loading conditions of $\bar{V} = 6$ $m/s$ with $TI = 20\%$ and $\bar{V} = 9$ $m/s$ with $TI = 15\%$ is discussed. Also, the effectiveness of the 2d-NTMDI in reducing the fatigue damage of wind turbine blades is discussed in this section. 

\subsection{Time-domain Response Mitigation Results}

In this subsection, the performance of the proposed 2d-NTMDI will be evaluated in reducing the bidirectional response of the blade. The optimized parameters obtained in the previous section will be used. As previously mentioned, the damper is located at $r_0 = 45$ $(m)$ from the blade root. Two different loading conditions of $\bar{V} = 6$ $m/s$ with $TI = 20\%$ and $\bar{V} = 9$ $m/s$ with $TI = 15\%$ are used.

\begin{figure}[ht]
	\centering
  \includegraphics[scale=0.3, trim= 0cm 6cm 0cm 4cm, clip=true]{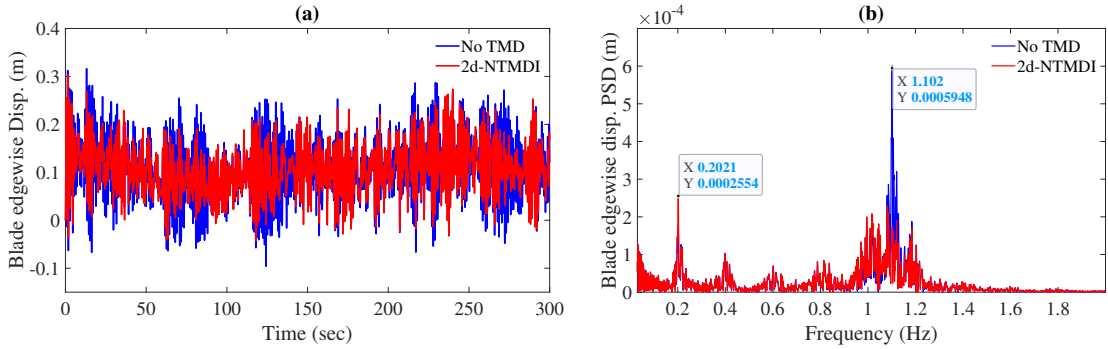}
	\caption{Edgewise displacement comparison between controlled and uncontrolled blades under $6$ $m/s$ mean wind speed: (a) edgewise response time-history, (b) edgewise response spectrum }
	\label{Edge_V6}
\end{figure}

Fig. \ref{Edge_V6} (a) illustrates the edgewise displacement time-history comparison between controlled and uncontrolled blades under $6$ $m/s$ mean wind speed and a $20\%$ turbulence intensity. As a representative case, a $6\%$ mass ratio is chosen with the optimized parameters: $\beta_x = 5\%$, $\beta_y = 80\%$, $\zeta = 17\%$, $f_{x,opt} = 1.12$ and $f_{y,opt} = 1.00$. Through comparison, one can clearly observe that the response of the blade in edgewise direction can be significantly mitigated using the 2d-NTMDI. Quantitatively, the RMS response of the blade in edgewise direction can be reduced by $18\%$. Detailed reduction effects of different mass ratios are listed in Table \ref{Tab_mu_V6}. Fig. \ref{Edge_V6} (b) illustrates the response spectrum of the data presented in Fig. \ref{Edge_V6} (a).  It can be seen that the spectral peak at $1.1$ $Hz$ which is the fundamental natural frequency of the blade in edgewise direction is significantly mitigated by the 2d-NTMDI. It should be noted that in Fig. \ref{Edge_V6} (b), the lower frequency peak at $0.2$ $Hz$ is caused by the rotational frequency of the blade.

\begin{figure}[ht]
	\centering
  \includegraphics[scale=0.3, trim= 0cm 6cm 0cm 4cm, clip=true]{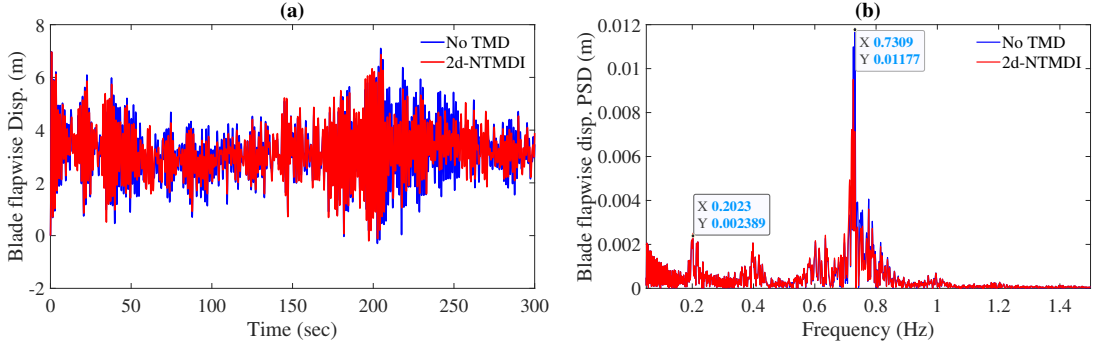}
	\caption{Flapwise displacement comparison between controlled and uncontrolled blades under $6$ $m/s$ wind speed: (a) flapwise response time-history, (b) flapwise response spectrum}
	\label{Flap_V6}
\end{figure}

Fig. \ref{Flap_V6} (a) illustrates the flapwise displacement time-history comparison between controlled and uncontrolled blades under $6$ $m/s$ mean wind speed and a $20\%$ turbulence intensity. It can be seen that the response of the blade in flapwise direction is mitigated using the 2d-NTMDI. Quantitatively, the RMS response of the blade in flapwise direction can be reduced by $11\%$. Detailed reduction effects of different mass ratios are listed in Table \ref{Tab_mu_V6}. Fig. \ref{Flap_V6} (b) illustrates the response spectrum of the data presented in Fig. \ref{Flap_V6} (a).  It can be seen that the spectral peak at $0.73$ $Hz$ which is the fundamental natural frequency of the blade in flapwise direction is  mitigated by the 2d-NTMDI. Similar to the response of the blade in edgewise direction (Fig. \ref{Edge_V6} (b)), the lower frequency peak at $0.2$ $Hz$ is caused by the rotational frequency of the blade.

\begin{figure}[ht]
	\centering
  \includegraphics[scale=0.3, trim= 0cm 6cm 0cm 4cm, clip=true]{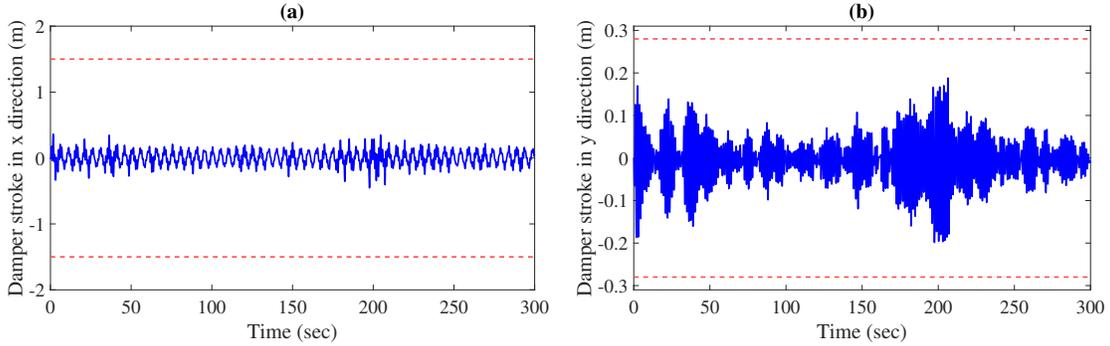}
	\caption{(a): 2d-NTMDI stroke in $x$ direction, (b): 2d-NTMDI stroke in $y$ direction. The dashed lines show the available space inside the blade in $x$ and $y$ directions. }
	\label{Blade_Stroke_V6}
\end{figure}

Figs. \ref{Blade_Stroke_V6} (a) and (b) illustrate the damper stroke in $x$ and $y$ direction under $6$ $m/s$ wind speed for a $6\%$ mass ratio. According to Fig. \ref{Blade_Stroke_V6}, (a) the damper stroke in $x$ direction is around $0.6$ $m$, while the available space inside the blade in this direction is $3$ $m$. Also, the damper stroke in $y$ direction is around $0.45$ $m$, while the available space inside the blade in this direction is $0.55$ $m$. Based on the available space inside the blade and the damper stroke, it can be concluded that the proposed 2d-NTMDI can be used inside the blade considering the space constraints.

\begin{center}
\begin{table*}[t]%
\centering
\caption{RMS response mitigation and damper stroke with different mass ratios under $\bar{V}=6$ $m/s$\label{Tab_mu_V6}}%
\begin{tabular*}{500pt}{@{\extracolsep\fill}lcccc@{\extracolsep\fill}}
\toprule
\textbf{Mass ratio} & \textbf{$RMS_{edge}$}  & \textbf{$RMS_{flap}$} & \textbf{$d_x (m)$} & \textbf{$d_y (m)$}  \\
\midrule
$2\%$    & $12.56\%$     & $3.02\%$   & $0.6$     & $0.43$ \\
$3\%$    & $15.88\%$     & $4.7\%$  & $0.74$     & $0.43$  \\
$4\%$    & $17.39\%$     & $5.88\%$  & $0.61$     & $0.37$  \\
$5\%$    & $19.14\%$     & $8.2\%$   & $0.72$     & $0.42$ \\
$6\%$    & $18.04\%$     & $10.5\%$   & $0.62$     & $0.47$ \\
$7\%$    & $19.14\%$     & $12.5\%$   & $0.64$     & $0.5$ \\
$8\%$    & $17.9\%$     & $14.5\%$   & $0.57$     & $0.53$ \\
\bottomrule
\end{tabular*}
\end{table*}
\end{center}

Table \ref{Tab_mu_V6} lists the RMS response mitigation and the damper stroke with different mass ratios ranging from $2\%$ to $8\%$ with the optimized parameters defined previously. One can find that by increasing the mass ratio, the effectiveness of the 2d-NTMDI increases in reducing the blade vibrations in edgewise and flapwise directions. Based on the results presented in Table \ref{Tab_mu_V6}, the 2d-NTMDI with a mass ratio of around $6\%$ can provide considerable mitigation and satisfies the limited space of the blade. A $6\%$ mass ratio has a physical mass of $85$ $kg$. Also, the optimized inertance ratio for $6\%$ mass ratio is $80\%$ which means a physical mass of $6$ $kg$. The damper stroke of the 2d-NTMDI for a $6\%$ mass ratio is $0.62$ $m$ in $x$ direction and $0.47$ $m$ in $y$ direction which is smaller than the available space inside the blade. Based on the above explanation, the proposed 2d-NTMDI can be used inside the blade practically to reduce the displacement of the blade in both directions.

\begin{figure}[ht]
	\centering
  \includegraphics[scale=0.3, trim= 0cm 6cm 0cm 4cm, clip=true]{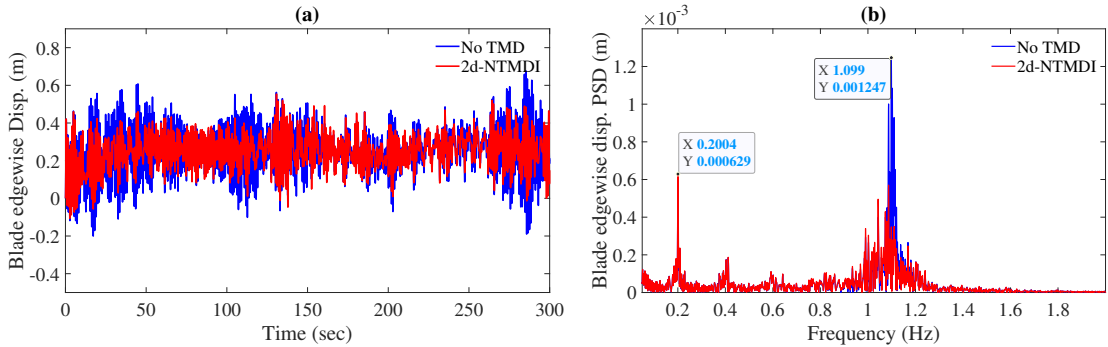}
	\caption{Edgewise displacement comparison between controlled and uncontrolled blades under $9$ $m/s$ wind speed: (a) edgewise response time-history, (b) edgewise response spectrum }
	\label{Edge_V9}
\end{figure}

Fig. \ref{Edge_V9} (a) illustrates the edgewise displacement time-history comparison between controlled and uncontrolled blades under $9$ $m/s$ wind speed and $15\%$ turbulence intensity. As a representative case, a $6\%$ mass ratio is chosen with the optimized parameters: $\beta_x = 5\%$, $\beta_y = 105\%$, $\zeta = 15\%$, $f_{x,opt} = 1.00$ and $f_{y,opt} = 1.00$. It can be observed that the response of the blade in edgewise direction can be considerably mitigated using the 2d-NTMDI. Quantitatively, the RMS response of the blade in edgewise direction can be reduced by $30\%$ with a $6\%$ mass ratio. Detailed reduction effects of different mass ratios under $9$ $m/s$ mean wind speed are listed in Table \ref{Tab_mu_V9}. Fig. \ref{Edge_V9} (b) illustrates the response spectrum of the data presented in Fig. \ref{Edge_V9} (a).

\begin{figure}[ht]
	\centering
  \includegraphics[scale=0.3, trim= 0cm 6cm 0cm 4cm, clip=true]{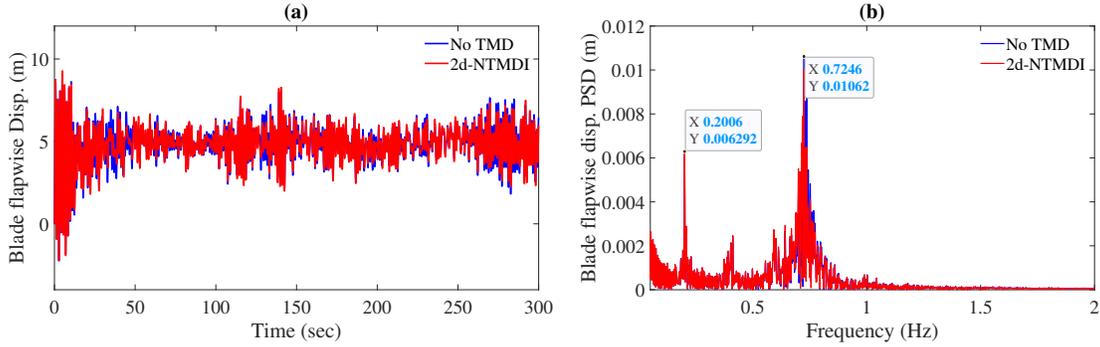}
	\caption{Flapwise displacement comparison between controlled and uncontrolled blades under $9$ $m/s$ mean wind speed: (a) flapwise response time-history, (b) flapwise response spectrum}
	\label{Flap_V9}
\end{figure}

Fig. \ref{Flap_V9} (a) illustrates flapwise displacement time-history comparison between controlled and uncontrolled blades under $9$ $m/s$ mean wind speed and a $15\%$ turbulence intensity. It can be seen that the effectiveness of the 2d-NTMDI in reducing the blade's response in flapwise direction is lower under $9$ $m/s$ mean wind speed than that of under $6 m/s$ mean wind speed. Quantitatively, the RMS response of the blade in flapwise direction can be reduced by $3\%$. Fig. \ref{Flap_V9} (b) illustrates the response spectrum of the data presented in Fig. \ref{Flap_V9} (a).

\begin{center}
\begin{table*}[t]%
\centering
\caption{RMS response mitigation and damper stroke of different mass ratios under $\bar{V}=9$ $m/s$\label{Tab_mu_V9}}%
\begin{tabular*}{500pt}{@{\extracolsep\fill}lcccc@{\extracolsep\fill}}
\toprule
\textbf{Mass ratio} & \textbf{$RMS_{edge}$}  & \textbf{$RMS_{flap}$} & \textbf{$d_x (m)$} & \textbf{$d_y (m)$}  \\
\midrule
$2\%$    & $15.18\%$     & $1\%$   & $0.89$     & $0.52$ \\
$3\%$    & $20\%$     & $1.5\%$  & $1.08$     & $0.5$  \\
$4\%$    & $24\%$     & $1.92\%$  & $1.13$     & $0.5$  \\
$5\%$    & $26\%$     & $2.5\%$   & $1.17$     & $0.54$ \\
$6\%$    & $29.7\%$     & $3\%$   & $1.24$     & $0.52$ \\
$7\%$    & $30\%$     & $3.1\%$   & $1.39$     & $0.55$ \\
$8\%$    & $30.1\%$     & $3.1\%$   & $1.15$     & $0.48$ \\
\bottomrule
\end{tabular*}
\end{table*}
\end{center}

Table \ref{Tab_mu_V9} lists the RMS response mitigation and the damper stroke with different mass ratios ranging from $2\%$ to $8\%$ with the optimized parameters defined previously. One can find that by increasing the mass ratio, the effectiveness of the 2d-NTMDI increases in reducing the blade vibrations in edgewise and flapwise directions. Based on the results presented in Table \ref{Tab_mu_V9}, a mass ratio of around $6\%$ can provide considerable mitigation in edgewise direction and satisfies the limited space of the blade. However, the mitigation effect in flapwise direction is weak. This is due to the reason that as the wind speed increases, the 2d-NTMDI needs a larger space to function. Yet, the limited space in the flapwise direction of the blade constraints the effectiveness of the proposed controller. Therefore, the proposed controller is more effective in reducing the vibrations of the blade in edgewise direction under larger wind speeds ($> 9 m/s$). Under smaller wind speeds ($< 9 m/s$), the 2d-NTMDI can provide effective response reduction for both directions.

\subsection{Fatigue damage mitigation}

\begin{figure}[ht]
	\centering
  \includegraphics[scale=0.3, trim= 5cm 5cm 10cm 0cm, clip=true]{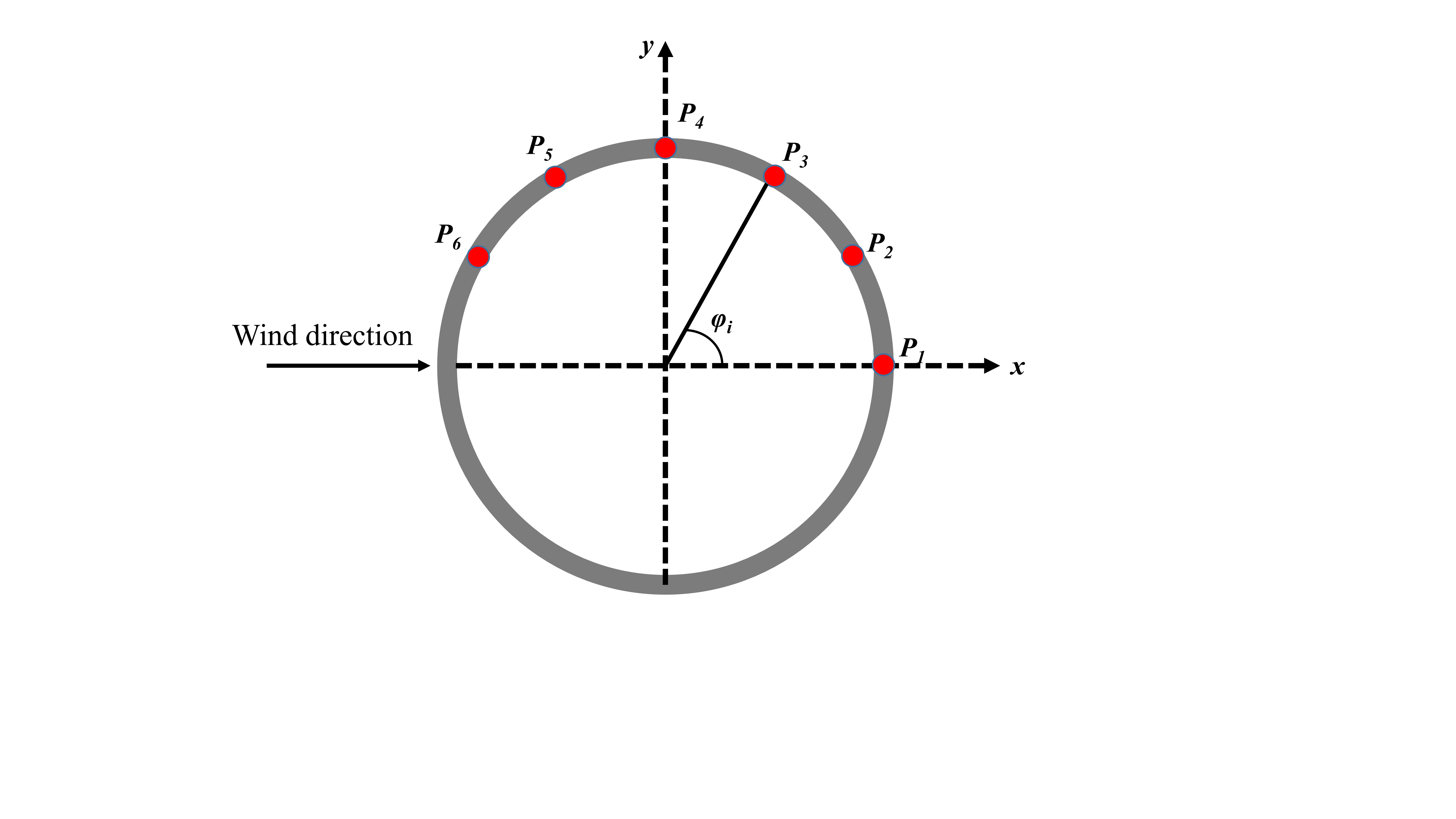}
	\caption{Representative points distributed along the out-most semi-circle of the blade root cross section}
	\label{fig_blade_root}
\end{figure}

The structural response of the blade in edgewise and flapwise directions results in bi-directional bending moment $M_e$, $M_f$ in the blades. The resultant stress distribution along the cross-section of the blade root becomes complicated because of the time-varying relative magnitude of the displacement in the edgewise and flapwise directions. Due to the symmetry at the blade root cross-section, $6$ representative points, which is referred to as $P_1$ to $P_6$. These points are evenly distributed along the out-most semi-circle of the blade root cross-section and are selected for fatigue damage calculation. Fig. \ref{fig_blade_root} demonstrates the chosen points. The resultant stress of these selected points is estimated as:

\begin{equation}
S_i = \frac{M_e}{I_e}c\sin{\phi_j} - \frac{M_f}{I_f}c\cos{\phi_j}
\label{eq10_15_1}
\end{equation}
where $M_e$ and $M_f$ denote the bending moments and $I_e$ and $I_f$ are the sectional moments of inertia of the blade root section.

\begin{figure}[ht]
	\centering
  \includegraphics[scale=0.3, trim= 0cm 0cm 5cm 0cm, clip=true]{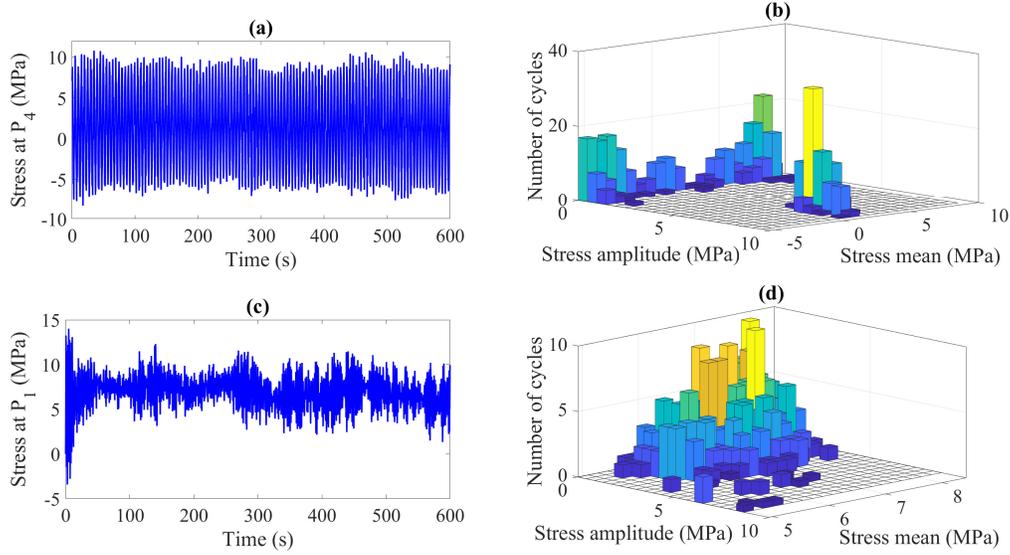}
	\caption{(a) Stress time-history at $P_4$, (b) rain-flow matrix of stress at $P_4$, (c) Stress time-history at $P_1$, (d) Rain-flow matrix of stress at $P_1$.} 
	\label{fig_RFC}
\end{figure}

Fig. \ref{fig_RFC} (a) and (c) portrays the stress time-history induced by the edgewise and flapwise vibrations, under a representative mean wind velocity of $9$ $m/s$. Fig. \ref{fig_RFC} (a) shows the stress of point $P_4$ caused by edgewsie vibration of the blade. Fig. \ref{fig_RFC} (c) shows the stress of point $P_4$ caused by flapwise vibration of the blade. The corresponding rain-flow matrix which demonstrates the stress amplitude, stress mean and the number of cycles is portrayed in Fig. \ref{fig_RFC} (b) and (d).

According to the calculated number of cycles and stress range, the cumulative fatigue damage of each point under each wind velocity is determined using the S-N curve and Miner's law. The cumulative fatigue damage of each point for both controlled and uncontrolled wind turbine blades is tabulated in Table \ref{Tab_fatigue_blade}. It should be noted that a representative mass ratio of $6\%$ is chosen to evaluate the performance of the 2d-NTMDI in reducing the fatigue damage of the blades. It can be found that the fatigue damage is larger at $P_3$ and $P_4$ than at the other points, which shows that the vibrations of the blade in edgewise direction dominate the fatigue damage development and can be the main cause of fatigue failure in blades. Considering the fatigue damage of the most dominant hot spot point $P_4$, one can find that the proposed 2d-NTMDI can improve the fatigue life by up to $35\%$.

\begin{center}
\begin{table*}[t]%
\centering
\caption{$600s$ cumulative fatigue damage with and without controller\label{Tab_fatigue_blade}}
\begin{tabular*}{500pt}{@{\extracolsep\fill}lcccc@{\extracolsep\fill}}
\toprule
\textbf{Location in outer circle} & \multicolumn{2}{c}{\textbf{Cumulative fatigue damage}} & \textbf{Reduction} \\ 
\midrule
                       &  No TMD                  & 2d-NTMDI                  &   \\ \hline
$P_1$                  &  $5.36 \times 10^{-15}$    &$5.23 \times 10^{-15}$       &$2.6\%$                           \\
$P_2$                  &  $3.77 \times 10^{-14}$    &$3.45 \times 10^{-14}$       &$8\%$                                                     \\
$P_3$                  &  $7.27 \times 10^{-14}$    &$5.39 \times 10^{-14}$       &$25\%$                                                     \\
$P_4$                  &  $8.2 \times 10^{-14}$     &$5.29 \times 10^{-14}$       &$35\%$                                                    \\
$P_5$                  &  $3.94 \times 10^{-14}$    &$2.37 \times 10^{-14}$       &$40\%$                                                    \\
$P_6$                  &  $8.05 \times 10^{-15}$    &$7.08 \times 10^{-15}$       &$12\%$                                          \\
\bottomrule 
\end{tabular*}
\end{table*}
\end{center}


\section{Conclusions}

The present paper proposes a novel two-dimensional nonlinear tuned mass damper inerter, to reduce the bi-directional vibrations of wind turbine blades. The proposed 2d-NTMDI has significant geometric nonlinearity due to the spatial arrangement of the springs, which complicates the optimum design. The optimum design parameters of the 2d-NTMDI are obtained using a numerical search method under different loading conditions. The performance of the 2d-NTMDI in mitigating the responses of the wind turbine blade in edgewise and flapwise directions is evaluated considering the limited space inside the blade. Also, the effectiveness of the 2d-NTMDI is examined in mitigating the fatigue damage of the wind turbine blades. According to the presented results and discussions, the following key conclusions can be obtained:

\begin{enumerate}

\item
A mathematical model of a novel two-dimensional nonlinear tuned mass damper inerter (2d-NTMDI) is derived for the first time. The optimum design parameters of the 2d-NTMDI are obtained using a numerical search approach for a wind turbine blade subjected to representative realistic wind loading. The optimum design parameters are obtained to minimize the RMS response of the wind turbine blade in edgewise and flapwise directions. 

\item
It is found that the 2d-NTMDI can effectively reduce the structural response of the wind turbine blade. Quantitatively, a $6\%$ mass ratio 2d-NTMDI can reduce the vibrations of the wind turbine blade in edgewise direction by around $18\%$, and in flapwise direction by around $11\%$ under $6$ $m/s$ mean wind speed. Under larger mean wind speeds such as $9$ $m/s$, the effectiveness of the 2d-NTMDI increases in edgewise direction, and decreases in flapwise direction. Quantitatively, a $6\%$ mass ratio 2d-NTMDI, can reduce the vibrations of the wind turbine blade in edgewise direction by around $30\%$, and in flapwise direction by around $3\%$ under $9$ $m/s$ mean wind speed.

\item
The proposed 2d-NTMDI can perform well inside the blade considering the limited space of the blade. This is beneficial for practical applications considering the limited space inside the blade. Also, the proposed 2d-NTMDI can be used in larger wind turbine blades such as DTU-10MW wind turbine and many other civil structures and mechanical systems. The available space inside the larger blades and other structures will allow the proposed 2d-NTMDI to experience a larger stroke, thereby providing better bi-directional mitigation. 

\item
It is found that the fatigue damage of points $P_3$ and $P_4$ of the out-most circle of the blade root cross-section is larger than any other location. This provides beneficial guidance for the design and analysis of the wind turbine blades. It is found that the proposed 2d-NTMDI can mitigate the fatigue damage of the blade root cross-section by up to $35\%$ under real wind conditions.

\item
Although the performance of proposed 2d-NTMDI is evaluated via a wind turbine blade, the proposed device has great potential to be applied in a large set of civil and mechanical structures, e.g., tall buildings and bridges that suffer from bi-directional vibrations.

\end{enumerate}

\section*{Acknowledgments}
This work was supported by Louisiana State Board of Regents Industrial Ties Research Sub-program ($AWD-001515$), USA. The authors are grateful for all the support.

\bibliographystyle{model1-num-names}

\end{document}